\documentstyle[sprocl]{article}

\input{psfig}
\def\be{\begin{equation}}
\def\ee{\end{equation}}
\def\bea{\begin{eqnarray}}
\def\eea{\end{eqnarray}}

\begin{document}

\title{THE STRUCTURE OF DARK MATTER HALOS. OBSERVATION VERSUS THEORY}

\author{ A. BURKERT}

\address{Max-Planck-Institut f\"ur Astronomie, K\"onigstuhl 17,\\
69117 Heidelberg, GERMANY}

\maketitle\abstracts{
The rotation curves of dark matter dominated dwarf galaxies are 
analysed. The observations show 
that dark matter halos represent a one-parameter family
with self similar density profiles. The global halo parameters, like
total mass and scale length are coupled by simple scaling relations.
The inner halo regions resemble non-singular, isothermal
spheres with constant density cores.
The observations are compared with dark matter halos,
resulting from cosmological cold dark matter simulations.
The theoretical models
predict that dark matter halos represent a one-parameter
family in agreement with the observations. 
However, in contradiction to the observations, the 
calculations lead to dark matter halos with $r^{-1}$ density cusps
in the center and non-isothermal velocity dispersion profiles.
Processes which might affect the inner halo structure,
resulting in isothermal, constant density cores are discussed.}
  
\section{Introduction}

Structure formation in the universe is strongly coupled with dark
matter (DM). Current cosmological models assume that there exists a
non-baryonic cold dark matter (CDM) component which consist of
non-relativistic particles that interact with the baryonic component
only through gravity. Given a primordial density fluctuation spectrum,
cosmological models investigate the formation of dark matter structures
and compare the results with observations of the distribution of galaxies
into clusters and superclusters~\cite{da}. 
This comparison provides important
information on cosmological parameters as well as on the initial
dark matter fluctuation spectrum and by this on the origin and nature of 
dark matter.

Dark matter can also be studied on galactic scales. On these
scales which are of order a few kpc, dark matter structures are in
general much older than their internal dynamical timescales. They
therefore have reached  a dynamical equilibrium state, a virialized 
dark matter halo. These halos could however still retain valuable
information about the initial conditions from which they formed,
if the assumption is valid, that dark matter consists of
collisionless and dissipationless particles.
Dark matter halos often host galaxies in their
inner regions. Studying the dynamical properties of these galaxies we 
gain insight into the inner density structure of dark matter halos 
and by this into the origin and nature of dark matter.

\section{The structure of simulated dark matter halos}
Assuming spherical symmetry, the radial DM mass distribution $M(r)$ can
be described by the DM rotation curve, that is the circular 
velocity profile $V_c(r) = (GM(r)/r)^{1/2}$.
The observations of constant circular velocities in the outer regions
of many spiral galaxies~\cite{cas} have lead to
the conclusion that DM halos are virialized isothermal spheres with an
$r^{-2}$ density profile in the observable radius range.
Cosmological models indeed lead to dark matter halos which produce 
constant outer rotation curves, in general agreement with these observations. 
Early cosmological calculations did not have enough resolution in order
to resolve the density structure of DM halos in detail. Recent high-resolution
simulations~\cite{n1,n2,n4} 
however have shown that in the inner and outer regions
dark matter halos depart significantly from an $r^{-2}$ power-law
distribution. All halo density profiles can be fit accurately
by the simple formula,

\begin{equation}
\rho (r) = \frac{\bar{\rho}}{(r/r_s)(1+r/r_s)^2}
\end{equation}

\noindent where $\bar{\rho}$ and $r_s$ are two free parameters.
It is very interesting that Navarro et al~\cite{n2,n4} find a strong 
correlation between $\bar{\rho}$ and $r_s$.
Dark matter halos seem to represent a one-parameter family,
characterized completely by their virial mass $M_{200}$ which is the total
mass inside the virial radius $r_{200}$. $r_{200}$ is the characteristic
radius inside which the mean DM density is $200 \times \rho_{crit}$,
where $\rho_{crit} = 3 H^2/8 \pi G$ is the critical density.

\section{The structure of observed dark matter halos}

Unfortunately it is difficult to observationally
verify these numerical results as galaxies are in general
gravitationally dominated by their visible baryonic components in
the inner regions, while in the outer regions there is not
enough visible material in order to measure accurately a rotation curve.
In the inner region the inferred DM profiles will depend
strongly on how much baryonic mass is subtracted, which in turn
depends on the assumed baryonic mass-to-light ratio. The situation
becomes even more complicated by the fact that a dominating baryonic
component will gravitationally affect and change the cold dark matter
density profile~\cite{n3}.

\begin{figure}
\psfig{figure=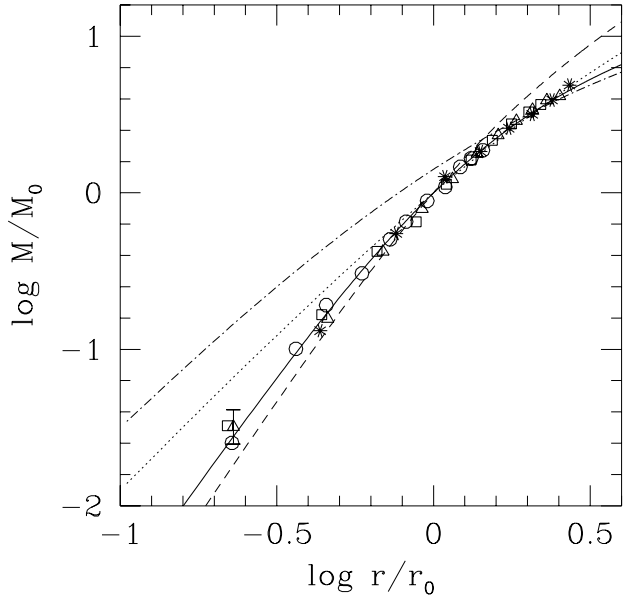,height=9.5cm}
\caption{Dark matter mass profiles are shown for the following
dwarf spiral galaxies:
DDO154 (open triangle),  DDO105 (open square),
NGC3109 (open circle) and DDO170 (starred). The errorbar at the
innermost triangle represents the observational uncertainty in the
inner region.  The isothermal fit with core radius $r_0$ is shown as
dashed curve, the solid line shows  the revised profile, given in the
text.  The dotted and dot-dashed curve show the mass profiles as
predicted from CDM calculations with formation redshifts of
$z=0.6$ and $z=1.5$, respectively.}
\end{figure}

\subsection{Observed dark matter mass profiles}

This situation has changed with the discovery of
a new class of low surface brightness dwarf spirals 
and irregulars, which are strongly
dominated by dark matter, even in their innermost regions. High-quality
rotation curves have become available in the past few years~\cite{br,ca,pu,puc},
which provide insight into the detailed structure of dark matter halos.

Figure 1 shows the dark matter mass profiles of four dwarf galaxies
with high signal-to-noise ratio HI rotation curves. All four profiles
indeed follow  the same universal mass relation. The dot-dashed and dotted
curves show DM profiles as predicted from cosmological simulations
(equation 1). The dot-dashed curve corresponds to a dark matter halo
with virial radius $r_{200}=5 \times r_s$.
The best fit through the data. using equation 1. is achieved with the dotted 
line which assumes $r_{200}=17.5 \times r_s$. 
Clearly, the halo profiles resulting from numerical simulations,
are too massive at small radii when compared with the observations.
This is a result of the central $r^{-1}$ density cusp. The apparent
contradiction between observation and theory has been discussed
in detail by Flores and Primack~\cite{fl}, Moore~\cite{mo} and Burkert~\cite{bu,bur}.

A nice fit through the observed profiles
over the whole observable radius range is achieved with the
simple density distribution~\cite{bu,bur},

\begin{equation}
\rho_{DM}(r) = \frac{\rho_0 r_0^3}{(r+r_0)(r^2+r_0^2)}
\end{equation}

\noindent where $\rho_0$ and $r_0$ are free parameters which represent
the central DM density and a scale radius, respectively.
Equation 2 resembles an isothermal profile with a constant-density core
at small radii ($r<r_0$). At large radii the density
decreases faster than expected for an isothermal distribution,
in agreement with the predictions from CDM calculations.

\subsection{Dark matter scaling relations}

Navarro et al~\cite{n2,n4} have shown that the two scale parameters
of equation 1 are strongly correlated. Small halos
are significantly denser than large halos as a result of
the fact that small, low-mass halos formed at higher collapse redshifts when the
density of the universe was higher. 

\begin{figure}
\psfig{figure=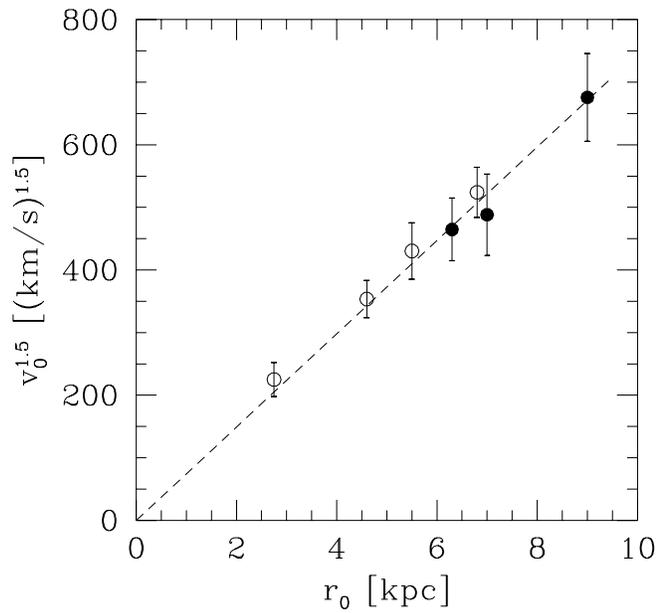,height=9.5cm}
\caption{The scaling relation between the rotational velocity $v_0$
measured at $r_0$ is shown for the
DM halos, investigated by Burkert$^2$. Open circles represent
the four DDO galaxies which have been used also in Fig. 1.
The filled circles show three additional galaxies: NGC55, NGC300 and
NGC1560. The dashed line is a fit through the data points.}
\end{figure}

Whether DM halos indeed represent a one-parameter family, being
described completely by their total mass, can be investigated
by looking for a correlation between the free parameters 
$\rho_0$ and $r_0$ in the observational fit formula (equation 2).
Instead of using $\rho_0$, which cannot be observed directly, Fig. 2
shows the rotational velocity $v_0$ of observed DM rotation curves
at $r_0$ as a function of $r_0$. $r_0$ is determined by fitting
the data with a velocity curve as predicted by equation 2. We find 
indeed a very strong correlation between $r_0$ and $v_0$. The slope agrees
well with the predictions from cosmological models~\cite{bu}.
Using equation 2 and assuming spherical symmetry one can derive
the following scaling relations for the observed DM halos:

\begin{eqnarray}
   v_0 & = & 17.7 \left( \frac{r_0}{kpc} \right)^{2/3}
   \frac{km}{s} \nonumber \\
   M_0 & = & 7.2 \times 10^7 \left( \frac{r_0}{kpc} \right)^{7/3}
   M_{\odot} \\
   \rho_0 & = & 2.7 \times 10^{-2} \left( \frac{r_0}{kpc} \right)^{-2/3}
   \frac{M_{\odot}}{pc^3}  \nonumber
\end{eqnarray}
 
\noindent where $M_0$ is the total dark matter mass inside $r_0$.
These relations indicate that dark matter halos indeed represent
a one-parameter family, in agreement with cosmological models.
 
\section{ On the origin of isothermal dark matter cores}

As shown in the last section, the shape of the cosmologically
predicted universal dark matter density profiles disagrees with the observations.
Whereas the observations indicate isothermal dark matter cores 
with constant density $\rho_0$ and constant velocity
dispersion $\sigma_0$, the cosmological models lead to cuspy cores with
density profiles $\rho \sim r^{-1}$ and velocity
dispersion profiles $\sigma (r) \sim r$. The simulated dark matter cores
are dynamically cold and dense. The observed dark matter cores are hotter and
less dense. In order to explain this difference
a mechanism has to be found which heats dark matter cores, increasing
their velocity dispersion and by this decreasing the central dark matter
density. 

\subsection{Cosmological initial conditions}
 
Navarro et al~\cite{n4} have investigated in detail
how the structure of DM halos depends on the
adopted cosmological model.
They find that the profiles are always well fitted by equation 1,
independent of halo mass, of the adopted initial density fluctuation spectrum,
and of the values of the cosmological parameters. Thus the problem cannot
be solved by selecting a certain cosmological model. 

This result is not surprising. It is well known that the violent
gravitational relaxation of collisionless particle systems leads to universal
equilibrium profiles, independent of the initial conditions~\cite{du}.
The final profiles of such systems
can be well described by a Hernquist profile~\cite{he},

\begin{equation}
\rho_h(r) = \frac{M}{2 \pi} \frac{a}{r(r+a)^3}
\end{equation}

\noindent where M is the total mass of the system and a is 
its scale length. The
Hernquist profile, for example, gives a good description of the surface
brightness profiles of elliptical galaxies, collisionless
stellar systems which have gone
through a stage of violent relaxation. Note, that the simulated halo profiles
(equation 1) are very similar to the profiles described
by equation (3), with the main difference being
a less steeply decreasing density distribution in the outermost regions.
This results from the fact that the Hernquist profile describes
systems with finite mass, whereas in cosmological models with $\Omega = 1$
halos will always accrete dark matter, leading to a mass profile that should
diverge logarithmically for large radii.

\subsection{Warm dark matter}

Dark matter is assumed to consist of collisionless particles, which interact
only by gravity. In this case, the 6-dimensional, microscopic
phase space distribution 
function (DF) $f(\vec{x},\vec{v})$ is a conserved quantity. The cuspy 
dark matter cores with $\rho \sim r^{-1}$ and $\sigma \sim r$
are characterized by  a DF which diverges as
$f \sim \rho/\sigma^3 \sim r^{-4}$. Given a critical phase space
density $f_{crit}$, there always exists a finite radius $r_{crit}$, inside
which $f > f_{crit}$. If the maximum phase space density of dark matter
would be finite ($f < f_{crit}$), the dark matter density profile should flatten
inside $r_{crit}$.

CDM particles formed with negligible initial velocity dispersion and therefore
with an infinitely large $f_{crit}$. No phase space limitations are imposed 
on CDM cores. The situation is different in the case
of warm dark matter, which starts with a finite initial velocity dispersion
and therefore with a finite $f_{crit}$. In this case, DM cores
might become isothermal inside
$r_{crit}$, where f approaches a constant and universal value
$f_{crit}$, that is determined by the initial dark matter temperature.
This idea can be tested. According to the equations 3, the central phase space density
of observed dark matter cores scales as $f_0 \sim \rho_0 / v_0^3 \sim r_0^{-8/3}$.
It decreases steeply with increasing core radius. The centers
of DM halos are not limited by a universal and finite maximum phase space density
which rules out warm dark matter as origin for isothermal dark matter cores.

\subsection {Secular dynamical processes}

As the problem cannot be solved by varying the initial conditions or
the nature of dark matter we 
have to focus on secular dynamical processes that might affect the  central
parts of the dark matter halos, after the halo formation phase.

Navarro et al~\cite{n3} have proposed a scenario,
where a gaseous disk forms in the centers of dark matter halos.
The disk potential dominates the central gravitational potential. 
The authors assume that after a vigorous
episode of star formation a large fraction of the total baryonic
component is expelled from the galaxy through supernova-driven winds.
They show, that a sudden loss of a large fraction of the 
total gravitational mass from the inner region would result in an
expansion of the dark matter core, decreasing the central DM density.
This scenario seems at first very
attractive. The observed scaling relations for dark matter cores
and the fact that dark matter halo profiles are universal would however
require significant fine tuning between the early cosmological collapse phase
and the secular energetic processes. It is unlikely that
DM halos would have self-similar density profiles if their inner
structure is subsequently changed by dynamical processes which
are not related to the collisionless
relaxation process which determined the outer DM profiles. In this
case, we would expect that dark matter halos are described by two
independent parameters. The first parameter determines their inner structure.
It will depend on the violence of the secular processes. The second
parameter determines the outer DM structure and depends on the 
cosmological merging history of the halo.

Substantial mass loss also seems unlikely in the case of the
DDO154~\cite{ca,puc}, a dwarf spiral galaxy with 
a dark matter halo of total mass $M_{DM} \approx 3 \times 10^9 M_{\odot}$,
containing an extended HI disk with total mass $2.5 \times 10^8 M_{\odot}$.
Navarro et al~\cite{n3} estimate that, prior to the mass loss epoch, the mass
of the gaseous disk should be of order 6 per cent of the total dark
matter mass. This ratio is in agreement with the disk-to-halo
mass ratio in DDO154, demonstrating that no substantial mass loss
has yet occured in this system. On the other hand, the rotation curve
of DDO154 clearly shows that its dark matter halo has an isothermal
DM core, which must have formed by a different mechanism.

\section{Summary}

The observations indicate that dark matter halos are self-similar, being
described completely by one free parameter. This surprising
universality makes it unlikely that the isothermal cores of
observed DM halos result from secular processes. 
The observed shallow central density profiles probably
formed as a direct result of the same processes, which lead
to the dark matter halos in the first place.
As discussed above, DM calculations do not produce 
isothermal halo cores. We therefore have to
conclude that some important, yet unknown physical features, related
to the nature and origin of dark matter, are still missing
in cosmological models.


\section*{References}
\begin{thebibliography}{99}
\bibitem{br} A.H. Broeils, {\it Astron. Astrophys.}, {\bf 256}, 19 (1992).
\bibitem{bu} A. Burkert, {\it Astrophys. J.}, {\bf 447}, L25 (1995).
\bibitem{bur} A. Burkert, in {\it IAU 171: New Light on Galaxy Evolution}, 
eds. R. Bender and R.L. Davies (Dordrecht; Kluwer), 175 (1996).
\bibitem{ca} C. Carignan and K.C. Freeman, {\it Astrophys. J.}, {\bf 332}, L33 (1988).
\bibitem{cas} S. Casertano and J.H. van Gorkom. 1991, {\it Astron. J.}, {\bf 101}, 1231 (1991).
\bibitem{da} M. Davis {\it et al}, {\it Astrophys. J.}, {\bf 292}, 371 (1985).
\bibitem{du} J. Dubinski and R. Carlberg, {\it Astrophys. J.}, {\bf 378}, 496 (1991).
\bibitem{fl} R.A. Flores and J.R. Primack, {\it Astrophys. J.}, {\bf 427}, L1 (1994).
\bibitem{he} L. Hernquist, {\it Astrophys. J.}, {\bf 356}, 359 (1990).
\bibitem{mo} B. Moore, {\it Nature}, {\bf 370}, 629 (1994).
\bibitem{n1} J.F. Navarro {\it et al}, {\it MNRAS}, {\bf 275}, 56 (1995).
\bibitem{n2} J.F. Navarro {\it et al}, {\it Astrophys. J.}, {\bf 462}, 563 (1996).
\bibitem{n3} J.F. Navarro {\it et al}, {\it MNRAS}, {\bf 283}, L72 (1996).
\bibitem{n4} J.F. Navarro {\it et al}, {\it astro-ph/9611107}, in press, (1997).
\bibitem{pu} D. Puche and C. Carignan, C., {\it Astrophys. J.}, {\bf 378}, 487 (1991).
\bibitem{puc} C.R. Purton and C. Carignan, C., {\it Bul. AAS}, {\bf 28/4}, 1320 (1996).
 
\end{thebibliography}
\end{document}